# A sudden period of high activity from repeating Fast Radio Burst 20201124A

Adam E. Lanman,[1,2] Bridget C. Andersen,[1,2] Pragya Chawla,[1,2] Alexander Josephy,[1,2]
Gavin Noble,[3,4] Victoria M. Kaspi,[1,2] Kevin Bandura,[5,6] Mohit Bhardwaj,[1,2] P. J. Boyle,[1,2]
Charanjot Brar,[1,2] Daniela Breitman,[7,3,4] Tomas Cassanelli,[3,4] Fengqiu (Adam) Dong,[8]
Emmanuel Fonseca,[1,2,9,6] B. M. Gaensler,[3,4] Deborah Good,[8,10,11] Jane Kaczmarek,[12]
Calvin Leung,[13,14] Kiyoshi W. Masui,[13,14] B. W. Meyers,[8] Cherry Ng,[3] Chitrang Patel,[1,3]
Aaron B. Pearlman,[1,2,*] Emily Petroff,[1,2,15,†] Ziggy Pleunis,[1,2]
Masoud Rafiei-Ravandi,[16,17] Mubdi Rahman,[18] Pranav Sanghavi,[5,6] Paul Scholz,[3]
Kaitlyn Shin,[13,14] Ingrid Stairs,[8] Shriharsh Tendulkar,[19,20] and Andrew Zwaniga[1,2]

[1]Department of Physics, McGill University, 3600 rue University, Montréal, QC H3A 2T8, Canada
[2]McGill Space Institute, McGill University, 3550 rue University, Montréal, QC H3A 2A7, Canada
[3]Dunlap Institute for Astronomy & Astrophysics, University of Toronto, 50 St. George Street, Toronto, ON M5S 3H4, Canada
[4]David A. Dunlap Department of Astronomy & Astrophysics, University of Toronto, 50 St. George Street, Toronto, ON M5S 3H4, Canada
[5]Lane Department of Computer Science and Electrical Engineering, 1220 Evansdale Drive, PO Box 6109, Morgantown, WV 26506, USA
[6]Center for Gravitational Waves and Cosmology, West Virginia University, Chestnut Ridge Research Building, Morgantown, WV 26505, USA
[7]Department of Physics, University of Toronto, 60 St. George Street, Toronto, ON M5S 1A7, Canada
[8]Department of Physics and Astronomy, University of British Columbia, 6224 Agricultural Road, Vancouver, BC V6T 1Z1 Canada
[9]Department of Physics and Astronomy, West Virginia University, PO Box 6315, Morgantown, WV 26506, USA
[10]Center for Computational Astrophysics, Flatiron Institute, 162 Fifth Ave, New York, NY, 10010, USA
[11]Department of Physics, University of Connecticut, 196 Auditorium Road, U-3046, Storrs, CT 06269-3046, USA
[12]Dominion Radio Astrophysical Observatory, Herzberg Research Centre for Astronomy and Astrophysics, National Research Council Canada, PO Box 248, Penticton, BC V2A 6J9, Canada
[13]MIT Kavli Institute for Astrophysics and Space Research, Massachusetts Institute of Technology, 77 Massachusetts Ave, Cambridge, MA 02139, USA
[14]Department of Physics, Massachusetts Institute of Technology, 77 Massachusetts Ave, Cambridge, MA 02139, USA
[15]Anton Pannekoek Institute for Astronomy, University of Amsterdam, Science Park 904, 1098 XH Amsterdam, The Netherlands
[16]Perimeter Institute for Theoretical Physics, 31 Caroline Street N, Waterloo, ON N25 2YL, Canada
[17]Department of Physics and Astronomy, University of Waterloo, Waterloo, ON N2L 3G1, Canada
[18]Sidrat Research, PO Box 73527 RPO Wychwood, Toronto, ON M6C 4A7, Canada
[19]Department of Astronomy and Astrophysics, Tata Institute of Fundamental Research, Mumbai, 400005, India
[20]National Centre for Radio Astrophysics, Post Bag 3, Ganeshkhind, Pune, 411007, India

## ABSTRACT

Corresponding author: Adam E. Lanman
adam.lanman@mcgill.ca

The repeating FRB 20201124A was first discovered by CHIME/FRB in November of 2020, after which it was seen to repeat a few times over several months. It entered a period of high activity in April of 2021, at which time several observatories recorded tens to hundreds more bursts from the source. These follow-up observations enabled precise localization and host galaxy identification. In this paper, we report on the CHIME/FRB-detected bursts from FRB 20201124A, including their best-fit morphologies, fluences, and arrival times. The large exposure time of the CHIME/FRB telescope to the location of this source allows us to constrain its rates of activity. We analyze the repetition rates over different spans of time, constraining the rate prior to discovery to $< 3.4$ day$^{-1}$ (at $3\sigma$), and demonstrate significant change in the event rate following initial detection. Lastly, we perform a maximum-likelihood estimation of a power-law luminosity function, finding a best-fit index $\alpha = -4.6 \pm 1.3 \pm 0.6$, with a break at a fluence threshold of $F_{\min} \sim 16.6$ Jy ms, consistent with the fluence completeness limit of the observations. This index is consistent within uncertainties with those of other repeating FRBs for which it has been determined.

## 1. INTRODUCTION

Fast radio bursts (FRBs) are microsecond- to millisecond-duration bursts of radio emission with dispersion measures (DMs) that indicate extragalactic origin. The origins of such bursts are still unclear, but a variety of plausible mechanisms have been proposed (Platts et al. 2019). The observation that some FRBs repeat (Spitler et al. 2016; CHIME/FRB Collaboration et al. 2019a,b; Fonseca et al. 2020) rules out cataclysmic sources as the only FRB mechanism. The discovery of repeating FRBs has helped in efforts to identify FRB host galaxies (e.g. Tendulkar et al. 2017; Marcote et al. 2017; Prochaska et al. 2019; Heintz et al. 2020; Bhandari et al. 2021; Bhardwaj et al. 2021a,b). The recent CHIME/FRB catalog (CHIME/FRB Collaboration et al. 2021) demonstrates that repeating FRBs tend to have morphologies distinct from non-repeating FRBs, characterized by narrower bandwidths and larger temporal widths (Pleunis et al. 2021).

Some repeaters have shown non-Poissonian behavior. The first FRB identified to repeat, FRB 20121102A, has clustered bursts (Scholz et al. 2016; Spitler et al. 2016; Oppermann et al. 2018). Rajwade et al. (2020) suggested a 157-day periodicity in its burst times, which was later confirmed by Cruces et al. (2021). CHIME/FRB Collaboration et al. (2020) showed that repeater FRB 20180916 has a 16.35-day periodicity, with a $\sim$4-day activity window. Clustering of events in time may offer an important clue to the emission mechanism of repeating FRBs, or imply a changing local environment.

In this paper, we present observations of FRB 20201124A, which entered a highly active state towards the end of March 2021. Using the precise localization reported by the Very Large Array (VLA), we are able to report on total exposure to the source of FRB 20201124A and give strong constraints on its event rate, demonstrating significant change in event rate over time. Further, we can take into account the weighting of the synthesized beam in measuring burst fluences, and estimate the luminosity function.

### 1.1. *Observations of FRB 20201124A*

* McGill Space Institute (MSI) Fellow.
* FRQNT Postdoctoral Fellow.
† Veni Fellow.

The CHIME/FRB collaboration detected a burst from FRB 20201124A on 2020 November 24 at UTC 08:50:41.885952 with DM= $415.31 \pm 0.63$ pc cm$^{-3}$, which was observed to repeat about four minutes later with DM= $414.31 \pm 0.31$ pc cm$^{-3}$. Four subsequent bursts from the same location were detected sporadically over the next few months – two in 2020 December and two more between February and mid-March. Towards the end of 2021 March, FRB 20201124A entered a period of high activity, as can be seen in Figure 1 where its bursts' times of arrival are marked with vertical lines. In the last week of March, CHIME/FRB detected new bursts from it almost every day. This prompted the collaboration to issue an Astronomer's Telegram (Lanman & the CHIME/FRB Collaboration 2021) to encourage follow-up observations. As a result, several other observatories soon reported detections of repeat bursts from the same sky region at similar DMs (Kumar et al. 2021a,b; Xu et al. 2021). The full set of measured source locations, DMs, and numbers of detected bursts are listed in Table 1 along with the bandpasses and observation times of the various instruments.

We reported a localization to $\sim 14'$ precision using stacked intensity data from previous repeats (Lanman & the CHIME/FRB Collaboration 2021). Following our report, the Australia Square Kilometre Array Pathfinder (ASKAP) carried out observations of the provided location and reported two bursts (Kumar et al. 2021b,a). From baseband data, ASKAP obtained a localization to arcsecond precision (Day et al. 2021), and identified SDSS J050803.48+260338.0 (hereafter SDSSJ0508) as a likely host galaxy, with a photometric redshift $z = 0.08 \pm 0.02$ catalogued by Pan-STARRS (Beck et al. 2021; Day et al. 2021). Observations by the MMT Observatory later found a spectroscopic redshift of $z = 0.0979 \pm 0.0001$ for SDSSJ0508 (Fong et al. 2021).

The Five-hundred-meter Aperture Spherical Telescope (FAST) collaboration soon reported that they had observed over a hundred bursts from FRB 20201124A between 2021 April 1 and 6, and identified four candidate host galaxies within a 1 arcmin field with a preference for SDSSJ0508 (Xu et al. 2021). The Karl G. Jansky Very Large Array (VLA) also reported a burst and localization to $2''$ precision, coincident with SDSSJ0508 (Law et al. 2021). The FAST and VLA localizations were consistent but significantly offset from the ASKAP position by about 2 arcsec. ASKAP later found a localization consistent with others using low-band data, and identified a potential higher-band systematic causing the observed offset (Day et al. 2021).

The positions ascertained from VLA, FAST, and low-band ASKAP data were later corroborated by the European VLBI Network (EVN) using observations from April 10 from six EVN stations. The EVN results pinpointed the source to a precision of 4 mas (Marcote et al. 2021). The full analysis from EVN (Nimmo et al. 2021) revealed an offset of $\sim 1.3$ kpc of the FRB from the optical center of the host galaxy SDSSJ0508.

The upgraded Giant Metrewave Radio Telescope (uGMRT) carried out observations on 2021 April 4 for four hours, at the position reported by ASKAP (Wharton et al. 2021a; Marthi et al. 2021), in a bandpass of 550 – 750 MHz (within the CHIME band). From the 48 bursts they detected, they localized the source to within an arcsecond (Wharton et al. 2021b), in agreement with other reported positions. They also fitted a power-law luminosity function and constrained the total scattering time and DM. We compare our fitted results to theirs in § 5.

On 2021 April 9, the 100-m Effelsberg radio telescope carried out observations at 1.36 GHz and detected 20 bursts from FRB 20201124A (Hilmarsson et al. 2021). They detected circular polarization in one of them – a first for repeating FRBs. They argue that the polarization properties are consistent with known magnetospheric emission from pulsars. They also found, through structure-maximizing DM searches, significant downward-drifting substructure.

Further radio imaging observations uncovered a persistent radio source at the FRB 20201124A location, called PRS201124. The upgraded Giant Metrewave Radio Telescope (uGMRT) detected a persistent radio source at (J2000) RA $05^h08^m03^s.43$, Dec $+26°03'38''.5$ (uncertainty $\pm0.9$"), with a flux density of $0.7 \pm 0.1$ mJy at 650 MHz (Wharton et al. 2021a). VLA/realfast later reported observations of a persistent source at the same location at 3 and 9 GHz, with flux densities $0.34\pm0.03$ and $0.15 \pm 0.01$ mJy, respectively (Ricci et al. 2021). Persistent emission was not detected on milliarcsec scales in the EVN observations (Marcote et al. 2021). From VLA observations, Ravi et al. (2021) found that the spectral energy distribution of PRS201124 is consistent with star formation activity for the host galaxy. This was supported by analysis of uGMRT and VLA observations in Fong et al. (2021), and by multiwavelength observations reported in Piro et al. (2021).

It is noteworthy that FRB 20201124A was only recently detected by CHIME/FRB, despite significant prior exposure, and then became highly active shortly after its detection. TNS queries found no recorded transient events within a degree of the reported VLA position prior to the initial CHIME/FRB detection. Altogether, CHIME/FRB had a total of 41.42 hours of exposure to the position of FRB 20201124A prior to the initial detection.

The steady observing cadence of CHIME/FRB offers a unique opportunity to discover new repeating FRB sources, study the statistics of repeat events, and identify patterns of repetition and clustering. From the recent surge in detections, and lack of prior detections, we can demonstrate strongly non-Poissonian repetition from FRB 20201124A.

The following section provides a background discussion of CHIME/FRB's observations and sensitivity to FRB events. §3 describes describes our analysis to characterize each of the detected bursts from FRB 20201124A. In §4.1 we perform a statistical analysis of the changing event rate, demonstrating that FRB 20201124A changed in its event rate considerably over time. In §4.2 and §4.3, we discuss the observed morphologies of the repeat bursts and perform a power-law fit to the burst luminosity function. These observations are placed in the broader context of repeating FRBs in §5.

## 2. OBSERVATIONS

The CHIME/Fast Radio Burst experiment (CHIME/FRB; CHIME/FRB Collaboration et al. 2018) is a backend system of the Canadian Hydrogen Intensity Mapping Experiment (CHIME)[1] telescope at the Dominion Radio Astronomical Observatory (DRAO) near Penticton, British Columbia, Canada. The telescope comprises four 20-m × 100-m cylindrical paraboloid reflectors oriented North-South, each with 256 equi-spaced antennas along the cylinder axis. These antennas are made up of two orthogonal linear-polarization feeds sensitive to a bandpass of $400 - 800$ MHz.

The CHIME/FRB backend forms a grid of 1024 total intensity beams: 256 rows evenly spaced in $\sin\theta$ from 60° North to 60° South, each row containing four beams separated by about 4° from East to West (Ng et al. 2017). Beamwidths range from 0.5° to 0.25° over the bandpass. Intensity samples from each beam arrives at a 0.983 ms cadence over 16384 frequency channels, to reduce intra-channel dispersion smearing. RFI rejection, known source tagging, recording calibration data, and FRB candidate identification are done in a multi-stage process described in detail in CHIME/FRB Collaboration et al. (2018). Raw intensity data are saved to disk when the candidate event has a signal to noise ratio (S/N) greater than 10. The pipeline is also capable of triggering a dump of raw baseband (voltage) data from a 34-s buffer upon detection (Michilli et al. 2021). Baseband data were saved for four of the reported events, and will the subject of future analysis.

| Experiment | Observations UTC | Center MHz | BW MHz | RA/Dec J2000 | | DM pc cm$^{-3}$ | Bursts | Reference(s) |
|---|---|---|---|---|---|---|---|---|
| CHIME | See Fig 1 | 600 | 400 | $05^h08^m06^s \pm 6'$ | $+26°11' \pm 14'$ | 413.52(5) | 33 | [a] |
| ASKAP | Apr 1 and 2 | 864.5 | 336 | $05^h08^m03.662^s \pm 1''$ | $+26°03'39.82'' \pm 1''$ | 412(3) | 2 | [b,c,d] |
| FAST | Apr 1 to June 6 | 1250 | 400 | $05^h08^m03.4^s \pm 1.3'$ | $+26°03'33.3'' \pm 1.3'$ | — | >100 | [e] |
| VLA | Apr 6 to Apr 7 | 1500 | 1000 | $05^h08^m03.50^s \pm 2''$ | $+26°03'37.8'' \pm 2''$ | 420(10) | 1 | [f] |
| uGMRT | Apr 5 | 650 | 200 | $05^h08^m03.459^s \pm 0.5''$ | $+26°03'39.14'' \pm 0.25''$ | 410.8(5) | 48 | [g] |
| EVN | Apr 10 & Apr 19 | 1374 | 256* | $05^h08^m03.5077^s \pm 2$mas | $+26°03'38.504''2$ mas | 412.0(7) | 18 | [h, i] |
| Effelsberg | Apr 9 | 1360 | 300 | — | — | 411.6(6) | 20 | [j] |

**Table 1.** Observations released in Atels or subsequent papers. Data are missing in some fields where the publication did not include a fitted value. Note that the bandwidth (BW) of EVN varies depending on which antennas are included in the network (between 128 and 256 MHz). References: [a] Lanman & the CHIME/FRB Collaboration (2021) , [b] Kumar et al. (2021b), [c] Kumar et al. (2021a), [d] Day et al. (2021), [e] Xu et al. (2021), [f] Law et al. (2021), [g] Wharton et al. (2021b), [h] Marcote et al. (2021), [i] Nimmo et al. (2021), [j] Hilmarsson et al. (2021).

These beams are searched for FRBs in real time using a four-stage triggered software pipeline, with steps labelled L0, L1, L2/L3, and L4 (see CHIME/FRB Collaboration et al. 2018). L0 is the rechannelization and beamforming step in the X-engine. L1 does an initial RFI rejection and incoherent dedispersion. L2/L3 combine results from multiple beams to identify likely candidates, screen RFI, and tag known sources. Raw intensity data are saved to disk for any events with a signal to noise ratio (S/N) greater than 10. Metadata are stored by L4 while raw buffered intensity data are saved to disk by L1. The pipeline is also capable of triggering a dump of raw baseband (voltage) data from a 34-s buffer upon detection (Michilli et al. 2021). Baseband data were saved for four of the observed FRB 20201124A events, and will be analyzed in future work.

The realtime pipeline estimates approximate source positions via the method described in CHIME/FRB Collaboration et al. (2019a). Ratios among per-beam S/N values are fit to a model of the synthesized beams for a grid of sky locations and intrinsic spectra, a task which ultimately reduces to a $\chi^2$ minimization. Uncertainty in these localizations is typically on the scale of a synthesized beam width, $\sim 20'$, but can be larger for events detected in multiple beams. Repeater candidates are identified as bursts whose positions and DMs are consistent to within measurement uncertainties (CHIME/FRB Collaboration et al. 2019b).

## 2.1. Exposure and Sensitivity

We compute the CHIME/FRB exposure to the location of FRB 20201124A using a model of the CHIME/FRB synthesized beams. Exposure time per day is calculated as the time during which the source location was within the FWHM at 600 MHz of any of the synthesized beams that were active on that day. Individual beams can go offline if a CPU node in the FRB backend that manages them is shut down, which reduces total exposure. Of course, if the whole system is shut down during the source transit, the exposure time is zero. Transits that occurred when the system was not operating at nominal sensitivity are also excluded. The exposure per day across the full observation history of CHIME/FRB (excluding an early commissioning phase) is shown in the top half of Figure 1.

The root-mean-square (RMS) noise each day is characterized by analyzing the distribution of S/N values of observed Galactic radio pulsars, as described in Josephy et al. (2019) (see also Fonseca et al.

2020; CHIME/FRB Collaboration et al. 2021). These RMS noise values relative to the mean noise are shown in the lower half of Figure 1, with colors indicating the number of pulsars used to estimate the RMS noise. The daily variation in noise does not exceed ∼10%.

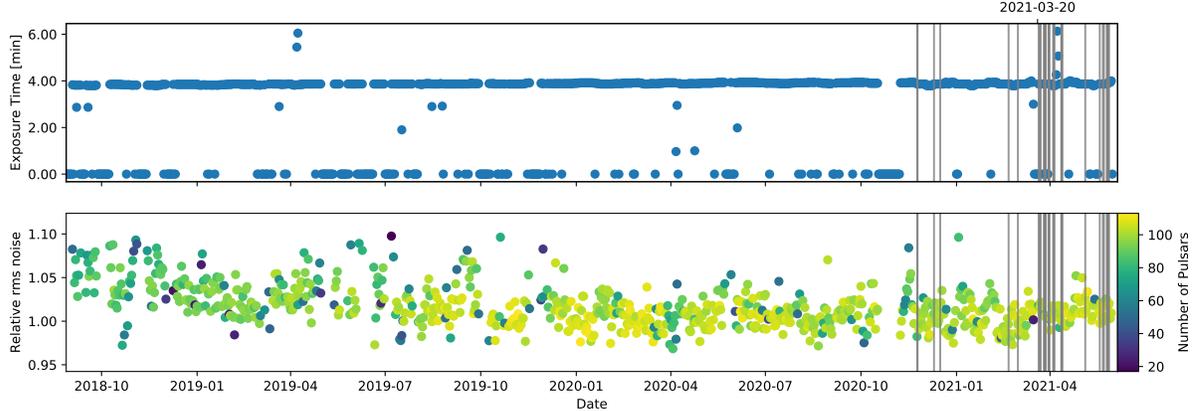

**Figure 1.** Exposure time per day (top) and relative rms noise (bottom) over time for the published position of FRB 20201124A. The location was within the 600-MHz FWHM of one or more synthesized beams of CHIME/FRB for approximately 4 mins each day. Thirty three vertical gray lines indicate times of FRB 20201124A repeat events. The date 2021 March 20 is indicated because this represents the approximate start of the period of high activity.

## 3. DATA REDUCTION

The intensity data of each event are processed in four analysis steps. The first searches for the DM that maximizes the signal to noise ratio (S/N) of the pulse integrated over frequency, as well as a structure-maximizing DM (see, e.g. Hessels et al. 2019). For the events reported here, the difference between structure-maximizing and S/N-maximizing DM was typically negligible, except for bursts with obvious substructure. The SNR-maximizing DM is used as an initial guess for the least-squares fitting routine `fitburst`, which fits a 2D analytic model to the two-dimensional dynamic spectrum (the full details of this model are discussed in CHIME/FRB Collaboration et al. (2021)). `fitburst` models each burst with one or more components $i$ and fits for the arrival time ($t_{\mathrm{arr},i}$), signal amplitude ($A$), scattering timescale ($\tau$) temporal width ($w_i$), power-law spectral index ($\gamma_i$), and spectral running ($r_i$). This model is fitted to the dynamic spectrum using an iterative $\chi^2$ minimization procedure. The DM and scattering timescale $\tau$ are fit globally regardless of the number of components (i.e. we assume the same DM and $\tau$ for all sub-bursts), and the scattering timescale is assumed to go as $\tau \propto \nu^{-4}$ with frequency $\nu$ (Lang 1971; Lorimer & Kramer 2005). If fitting for subbursts, the expected number, relative amplitudes, and separation in time of the subbursts are entered manually. These are initial guesses to help `fitburst` converge on the true values. Table 2 lists the best-fit measured properties of each detected burst, obtained from `fitburst`. Figure 2 shows waterfall plots for each of the events, along with the `fitburst` models for time series and spectrum.

For each event, `fitburst` is run automatically twice to determine the significance of scattering. The first iteration assumes a scattering timescale $\tau = 0$, such that the burst components have simple Gaussian shapes, while the second round fits for the scattering timescale as well. The $\chi^2$ values of the two fits are compared in an F-test to determine the significance of the scattering fit. We report scattering timescales in Table 2 only for bursts whose F-test results showed significant evidence of scattering ($p < 0.001$). We also report the best-fit DM of each burst, rather than dedisperse all events to the same DM, in order to look for evidence of a changing DM over time.

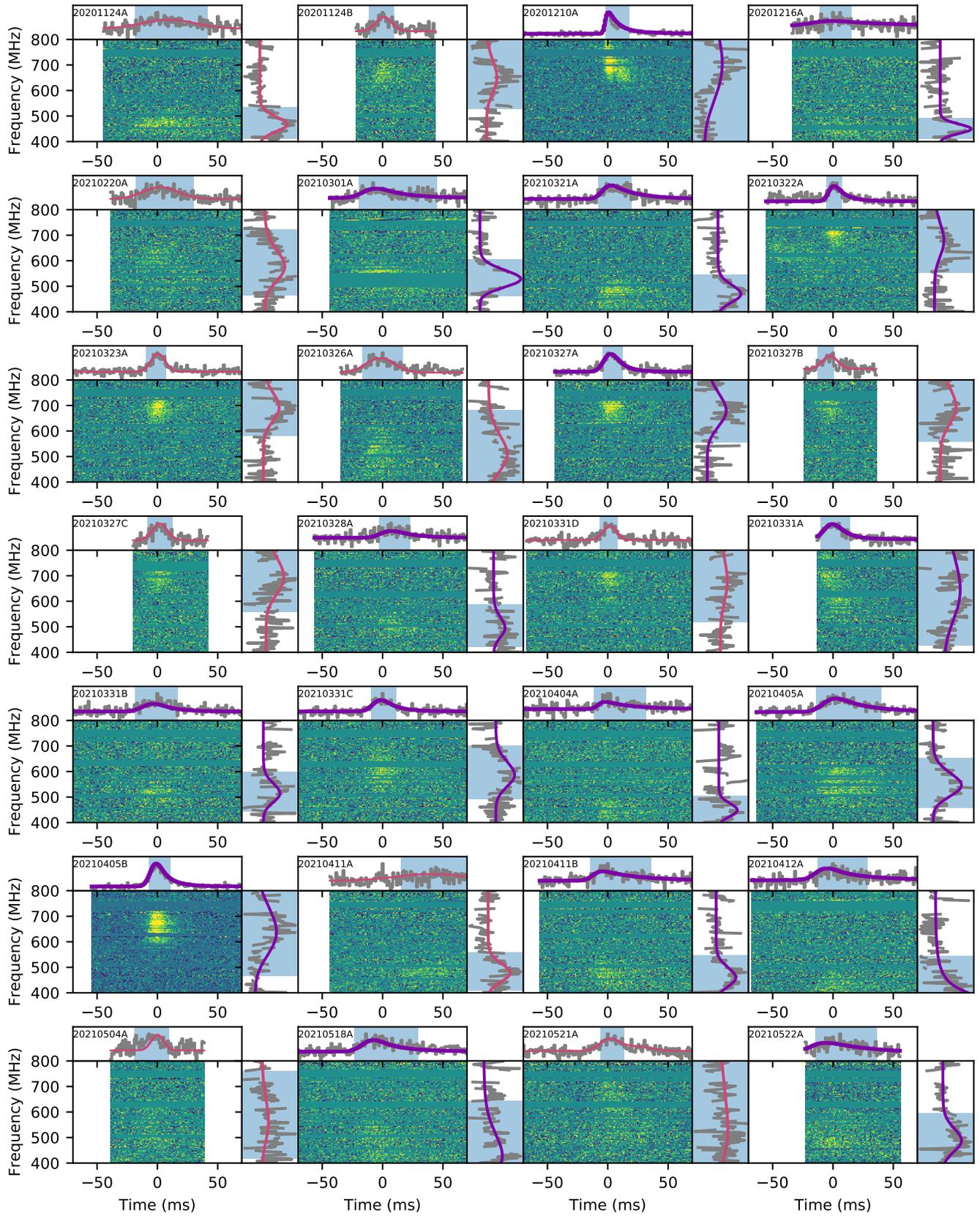

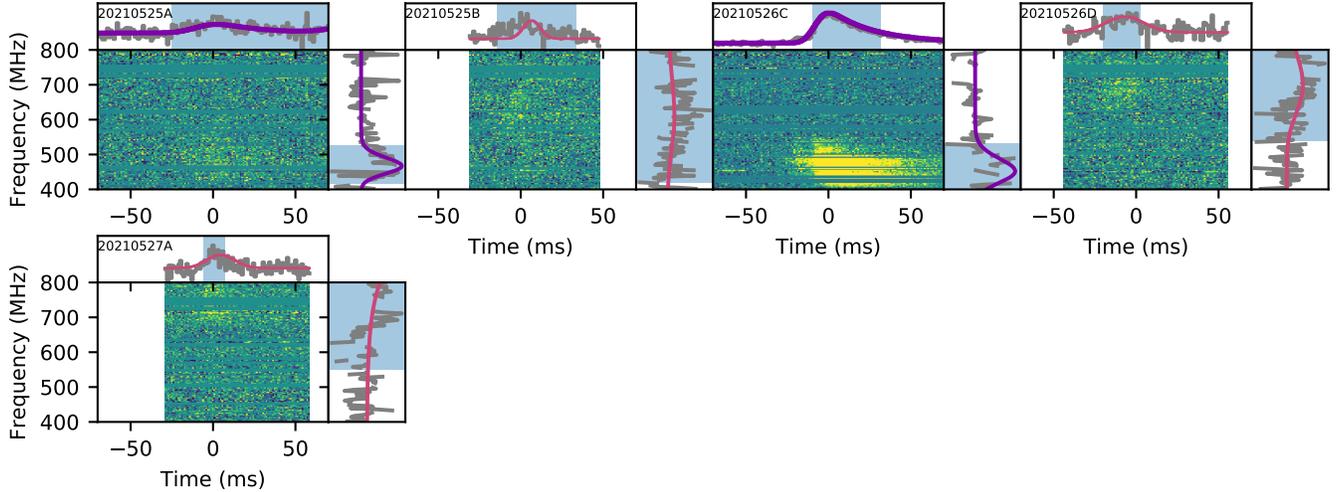

**Figure 2.** Waterfall plots for all events, shown with the same time window. These show only data included in the `fitburst` analysis, which takes a finite window around the time of arrival. For each, we plot the integrated spectrum on the right and integrated time series on top, along with the fitted time series and spectrum curves. The fitted time series and spectra are colored red when significant scattering is measured and purple otherwise.

For all but three of the FRB 20201124A bursts presented here, we assume a single component profile. The other three (20201210A, 20210521A, and 20210525A) could be confidently fit with a second, downward-drifting burst, suggesting that other bursts from this source could also have downward drifting components that `fitburst` could not resolve. The dynamic spectra for some of the brighter bursts show hints of subcomponents, but `fitburst` was unable to resolve them. Other repeating FRBs have been shown to have complex downward-drifting substructure, such as FRB 20181222 (CHIME/FRB Collaboration et al. 2019b) or FRB 20121102 (Gourdji et al. 2019). Recent results from the uGMRT show several FRB 20201124A bursts having downward-drifting substructure (Marthi et al. 2021). Downward drifting structure can be confused for a higher DM or scattering tail, and so some caution is needed when interpreting these results.

The intensity data are calibrated offline using the pipeline described by CHIME/FRB Collaboration et al. (2019a). Steady calibrator sources with known spectral properties are observed at meridian transit (i.e., at primary beam center). For each burst, the CHIME/FRB backend identifies the calibrator transit closest in declination and time, and then uses the measured spectrum to convert beamformer units to Jy. For all of the FRB 20201124A events, the calibrator was the Seyfert galaxy NGC 7720. The calibrated dynamic spectrum is then integrated over the bandpass and over the pulse width to estimate the total fluence. For sources of unknown position, there is an uncertainty in the difference in directional gain between the calibrator position and the FRB location. Normally, this pipeline assumes the source is at the meridian where the beam is strongest, and thus provides a lower bound on the fluence. With the localization provided by the interferometric measurements (Law et al. 2021) and a composite beam model, we are able to scale the calibrated flux estimates by the ratio of beam values between the calibrator and the true position of FRB 20201124A and report a flux estimate instead of a lower bound. However, this scaling also has the effect of amplifying noise for bursts detected toward the edges of the formed beam, so we only applied this scaling when the known source location was within the FWHM of the beam at 600 MHz. Flux/fluence values which are provided as only a lower bound are marked with a ">" in Table 2, and comprise 14 of the 33 total measurements.

| TNS Name | UTC Time of Arrival | DM pc cm$^{-3}$ | Width ms | Peak Flux Jy | Fluence Jy ms | Bandwidth MHz | S/N | Scattering Timescale ms |
|---|---|---|---|---|---|---|---|---|
| 20201124A | 2020-11-24 08:50:41.871 | 415.3(6) | 22(1) | 0.8(5) | 6(2) | 409.75 − 533.63 | 29.47 | − |
| 20201124B | 2020-11-24 08:54:45.982 | 414.1(3) | 7(2) | 0.6(4) | 5(2) | 529.28 − 800.20 | 17.07 | − |
| 20201210A | 2020-12-10 07:54:42.909, +33.00 ms | 411.1522(39) | 1.221(90), 0(7) | >2.6(8) | >26(5) | 447.99 − 800.20 | 72.30 | 14.2(5) |
| 20201216A | 2020-12-16 07:29:19.428 | 414.1(3) | 8(2) | >0.33(15) | >4.2(7) | 410.56 − 491.37 | 17.77 | 12(4) |
| 20210220A | 2021-02-20 03:06:22.937 | 420.3(4) | 14(2) | 0.9(6) | 5.8(9) | 463.78 − 720.79 | 22.00 | − |
| 20210301A | 2021-03-01 02:28:45.207 | 415.363(54) | 8.2(7) | >0.45(17) | >12(2) | 461.98 − 605.51 | 49.22 | 14(2) |
| 20210321A | 2021-03-21 01:12:22.535 | 412.5(2) | 4.1(6) | 0.8(4) | 9(3) | 402.94 − 547.30 | 32.36 | 8.3(7) |
| 20210322A | 2021-03-22 01:10:44.328 | 412.0(3) | 3.2(7) | 0.9(5) | 19(5) | 553.33 − 800.20 | 25.70 | 14(2) |
| 20210323A | 2021-03-23 01:08:04.091 | 413.5(2) | 5.9(7) | >0.7(2) | >8(2) | 582.85 − 800.20 | 38.78 | − |
| 20210326A | 2021-03-26 00:53:51.109 | 414.08(17) | 11.6(6) | >0.6(2) | >11(4) | 400.20 − 682.48 | 29.57 | − |
| 20210327A | 2021-03-27 00:46:22.284 | 417.4(2) | 4.3(8) | >0.5(2) | >8(2) | 555.07 − 800.20 | 36.89 | 12(2) |
| 20210327B | 2021-03-27 00:53:10.609 | 415.6(4) | 6.0(10) | 1.8(10) | 25(6) | 558.26 − 800.20 | 19.10 | − |
| 20210327C | 2021-03-27 00:53:10.641 | 417.5(4) | 6.1(10) | 1.4(8) | 21(5) | 559.44 − 800.20 | 19.22 | − |
| 20210328A | 2021-03-28 00:49:08.250 | 414.4(6) | 6(2) | 0.27(17) | 3.2(8) | 419.81 − 586.73 | 14.17 | 8(2) |
| 20210331D | 2021-03-31 00:33:06.545 | 416.4(3) | 5.5(4) | 1.2(6) | 14(4) | 518.49 − 800.20 | 21.04 | − |
| 20210331A | 2021-03-31 00:37:18.719 | 417.00(19) | 5.5(6) | >0.6(3) | >9(3) | 426.75 − 800.20 | 29.75 | 8(2) |
| 20210331B | 2021-03-31 00:38:34.938 | 412.6(5) | 5(1) | >0.5(2) | >6(2) | 442.54 − 599.30 | 18.81 | 8(1) |
| 20210331C | 2021-03-31 00:39:18.305 | 414.8(3) | 5.3(7) | >0.6(3) | >9(2) | 492.15 − 700.43 | 26.40 | 11(2) |
| 20210404A | 2021-04-04 00:41:43.318 | 410.1(3) | 3(2) | >0.32(15) | >5(1) | 400.20 − 505.53 | 17.22 | 8(1) |
| 20210405A | 2021-04-05 00:13:36.914 | 414.0(2) | 7(1) | 1.1(6) | 23(6) | 456.14 − 652.58 | 44.35 | 19(1) |
| 20210405B | 2021-04-05 00:13:36.664 | 413.279(16) | 3.9(2) | 2.3(9) | 30(9) | 465.10 − 800.20 | 82.25 | 12.2(8) |
| 20210411A | 2021-04-11 23:43:24.320 | 416(1) | 29(3) | >0.28(18) | >7(2) | 408.56 − 559.46 | 18.83 | − |
| 20210411B | 2021-04-11 23:49:52.839 | 411.75(14) | 5(1) | 1.0(5) | 17(6) | 400.20 − 546.50 | 21.83 | 12(2) |
| 20210412A | 2021-04-12 23:36:08.643 | 412.95(15) | 7(2) | >0.43(15) | >6(1) | 400.20 − 545.52 | 19.87 | 10(2) |
| 20210504A | 2021-05-04 22:18:21.634 | 413.844(50) | 5.1(7) | 1.8(10) | 11(3) | 415.95 − 759.68 | 10.69 | − |
| 20210518A | 2021-05-18 21:21:03.774 | 413.4(2) | 7(1) | 0.7(4) | 9(3) | 400.20 − 643.57 | 21.79 | 8(1) |
| 20210521A | 2021-05-21 21:07:19.468, +24.00 ms | 415.81(18) | 8(1), 12(5) | 1.5(7) | 17(4) | 400.20 − 800.20 | 19.65 | − |
| 20210522A | 2021-05-22 21:08:52.291 | 414.0(4) | 8(3) | 2(1) | 22(6) | 400.20 − 595.50 | 24.51 | 18(3) |
| 20210525A | 2021-05-25 20:53:21.903, +77.00 ms | 416.6(4) | 10(2), 9(3) | 0.7(4) | 19(6) | 417.66 − 527.28 | 29.65 | 10(3) |
| 20210525B | 2021-05-25 20:53:24.482 | 413.839(40) | 5.1(6) | 0.8(4) | 7(2) | 419.52 − 800.20 | 11.00 | − |
| 20210526C | 2021-05-26 20:45:38.653 | 410.783(12) | 5.48(11) | >3(1) | >140(47) | 400.20 − 800.20 | 301.17 | 10.42(12) |
| 20210526D | 2021-05-26 20:49:27.078 | 418.25(13) | 11(2) | 0.8(5) | 11(3) | 540.78 − 800.20 | 18.22 | − |
| 20210527A | 2021-05-27 20:44:49.565 | 418.663(58) | 8(1) | >0.7(3) | >7(2) | 550.96 − 800.20 | 14.39 | − |

**Table 2.** Properties of the repeat bursts from the source of FRB 20201124A detected by CHIME/FRB prior to 2021 May 27. Pulse bandwidth is defined as the full width at one tenth maximum (FWTM) of the pulse spectrum, as measured in the beam with the highest detection S/N. Scattering timescale is reported only for bursts which showed significant evidence of scattering. For the events with multiple components, times of arrival of later components are reported as offsets from the arrival time of the earliest component.

## 4. RESULTS

### 4.1. Temporal Study

Figure 1 shows that CHIME/FRB observed the FRB 20201124A source location for two years, with a total exposure time of about 1.73 days, prior to the first detected burst. By eye, it is clear that the repetition rate of FRB 20201124A is not constant and has changed substantially over the period of weeks. We can rule out constant-rate (Poissonian) repetition with a series of Kolmogorov-Smirnov (KS) tests on the times between detected events. For a Poisson process with rate $\lambda$, the time $t$ between events is exponentially-distributed, with probability density function (PDF) $f(t) = \lambda \exp(-\lambda t)$. However, CHIME/FRB only observes the source location for a limited time each sidereal day. For a periodically-sampled Poisson process, the total *exposure* time between bursts is exponentially distributed. The PDF thus has the form of an exponential of the total observed time, multipled by a periodic window,

$$f(t) = \int\limits_{0}^{a} \lambda e^{-\lambda \tau(t,T,a,t_a)} dt_a$$

$$\text{where } \tau(t,T,a,t_a) = \left\lfloor \frac{t+t_a}{T} \right\rfloor a + (t+t_a) \bmod T, \tag{1}$$

where $T = 23.93$ hr is the period between observing windows (one sidereal day), $a = 3.13$ min is the duration of the observation window, here taken to be the mean daily exposure time of CHIME/FRB. The brackets $\lfloor \cdot \rfloor$ denote the floor function. The probability of another event occuring at time $t$ after an initial event at $t_a$ is exponential in the total observed time $\tau$ after $t_a$. To get the probability of $t$, we marginalize over $t_a$ by integrating over a single window.

Non-detection over the pre-discovery total observed time rules out Poisson rates higher than 3.4 events per day at the $3\sigma$ level, for our sensitivity threshold, based on Equation (1). For times after the burst was discovered, we examine the distribution of inter-arrival times among three subsets of CHIME/FRB FRB 20201124A bursts:

- All events

- Bursts that arrived between 2020 Nov 24 and 2021 March 20

- Bursts that arrived after 2021 March 20.

The 2021 March 20 date was chosen as an approximate mark to the start of increased activity.For each subset, we apply a KS test to determine if the data are inconsistent with the distribution of Equation (1) for a range of possible event rates $\lambda$. The two-sided KS test measures the maximum distance between the empirical cumulative distribution function (ECDF) and the CDF corresponding with Equation (1). The null hypothesis of the KS test is that the data *are* drawn from the sampled exponential distribution.

Figure 3 shows the p-value of the KS test for a variety of event rates for the full set of events (blue), events before 2021 March 20, and events after 2021 March 20 (orange). The gray horizontal line marks $p = 0.05$, and the black horizontal line marks $p = 0.0027$, for $3\sigma$ significance. Over the full set of events, we can rule out constant Poissonian repetition to $p = 0.005$ significance (the maximum in the inset). For the sets of events before or after 2021 March 20 Poissonian repetition cannot be ruled out, but the acceptable event rates are inconsistent. For events after March 20, mean event rates smaller than 92 day$^{-1}$ and larger than 201 day$^{-1}$ can be ruled out to $3\sigma$ significance. To the same significance, rates smaller than 3.36 day$^{-1}$ and larger than 60 day$^{-1}$ can be ruled out for events between the initial detection and March 20.

The periodic sampling of CHIME/FRB can complicate the interpretation of these results. It is possible that FRB 20201124A could be repeating periodically, in such a way that for the first two years of CHIME/FRB observations it was always out of phase with its transits over the telescope. This periodicity would need to be very close to a multiple of a sidereal day for the FRB to escape detection until 2020 November 24. We note that the detections by ASKAP, VLA/realfast, FAST, and others did not correspond with the CHIME/FRB observation window, and so disfavor periodicity on a time scale near 24 hours. We also conducted searches for periodicity in the burst arrival times, which included H-tests (de Jager et al. 1989) and a search for peaks in the event time periodogram. These searches did not turn up any evidence of periodicity on scales between 1 day and 178.5 days.

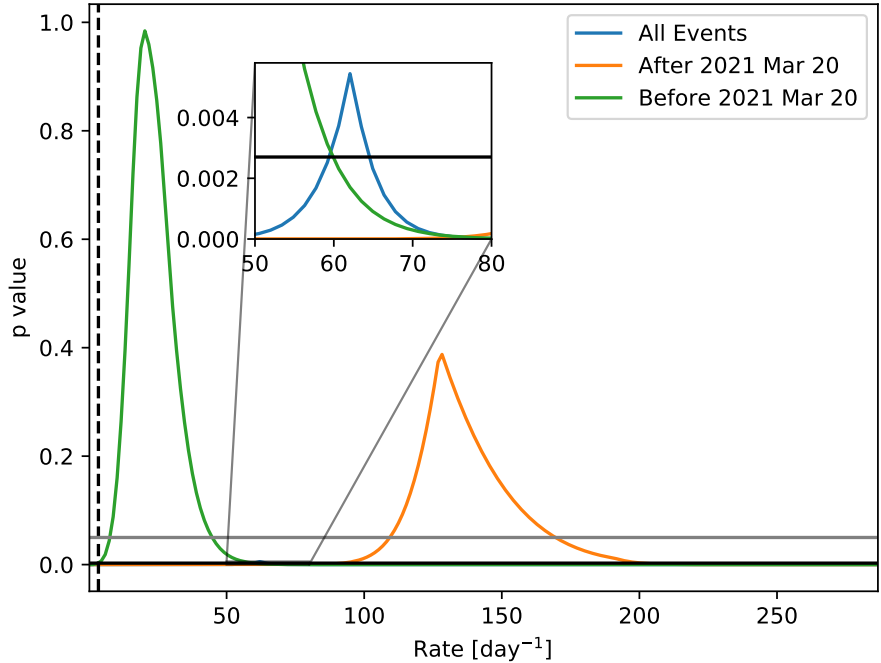

**Figure 3.** KS test p-value vs. event rate for two subsets of data. The horizontal black and gray lines mark p=0.0027 (i.e., $3\sigma$) and p=0.05, respectively. The vertical dashed line indicates the $3\sigma$ maximum event rate given non-detection prior to 2020 November 24. The blue curve gives the KS test results for all events following the initial detection. The orange curve is for events after 2021 March 20, our fiducial start date for the period of high activity. The green curve is for events prior to the start of high activity. The inset magnifies the region of event rates most consistent with all CHIME/FRB events. This shows that there is no single Poisson rate consistent with the repetition of FRB 20201124A.

We note that these results appear to be inconsistent with the burst rate of $\sim 16$ hour$^{-1}$ (384 day$^{-1}$) reported by uGMRT in Marthi et al. (2021). The uGMRT observations comprise 3 hours of exposure to the 20201124A location on the 2021 April 5. During this time, uGMRT found 48 bursts in the bandpass of $550 - 750$ MHz. The orange curve in Figure 3 tapers off sharply to zero at rates larger than about 210 day$^{-1}$, indicating that the sample of events after 2021 March 20 is inconsistent with high event rates. This apparent inconsistency suggests that the event rate changed during the period from 2021 March 20 to May 19.

## 4.2. *Burst Morphology*

Figure 4 shows the DM, pulse width, scattering timescale, and pulse bandpass of each burst vs. the time of detection, using the data in Table 2. We find no evidence of a consistent secular evolution in any of these parameters over time.

There is considerable apparent variation in DM and scattering timescale among these events. The highest S/N burst (20210526C) has a S/N-maximizing DM of $410.78 \pm 0.012$ pc cm$^{-3}$, a value consistent with DMs reported by other instruments in Table 1. The two bursts with the next highest S/N (20201210A and 20210405B) have slightly different scattering timescales ($14.24 \pm 0.55$ and $12.17 \pm 0.82$ ms), and DMs ($411.1522 \pm 0.0039$ and $413.279 \pm 0.016$ pc cm$^{-3}$). Of those two, the burst with the smaller DM (20201210A) also has a resolved subcomponent. If 20210405B has a downward-drifting subcomponent like 20201210A, the expected effect would be to confuse the second component for dispersion and scattering, which could explain the discrepancy. For the much wider and fainter bursts it is very likely that `fitburst` simply cannot resolve subcomponents and over-estimates the DM as a result. The biggest outlier is FRB 20210220A, which has a best-fit DM of $420.26 \pm 0.41$ pc cm$^{-3}$ and a relatively narrow bandpass and a large temporal width. The error bars in Figure 4 do not include the degeneracy between substructure and DM, and so should not be interpreted as showing significant DM variation.

The observed pulse bandpasses are shown in the bottom plot of Figure 4. These are unaffected by the confusion between unresolved sub-components and DM. There is a hint of bimodality in the peak frequencies, with several burst spectra clustering around $\sim 650$ MHz and others around or below 500 MHz. We plot the distribution of peak frequencies in a way analogous to Gaussian kernel density estimation with the estimator

$$\hat{g}(f_p) \propto \sum_i \exp\left(\frac{-(f_p - f_{pi})^2}{2\sigma_i^2}\right), \tag{2}$$

where $f_{pi}$ is the peak frequency of the $i^{\text{th}}$ burst, and the expression is normalized to unity. The uncertainty in each measured peak frequency is $\sigma_i = (0.233)\text{BW}_i$, where $\text{BW}_i$ is the bandwidth of the each burst and the factor of 0.233 scales the FWTM to standard deviation. Figure 5 shows the resulting distribution estimator for the FRB 20201124A data shown here, along with the same curves for the two most active repeaters in the first CHIME/FRB catalog (CHIME/FRB Collaboration et al. 2021). The peak frequencies of each burst are marked by the lines on the bottom. The distribution of FRB 20201124A peaks near 470 MHz and by 668 MHz. The other two similarly peak near 668 MHz. FRB 20180916B also looks to be bimodal, with a second peak near the bottom of the band.

There are several potential systematic causes of such a bimodality. If channels away from the peaks were more often contaminated with RFI, then it could lead to a preference for detecting bursts near the peaks. However, the distribution of RFI-excised channels shows no such pattern. Another possibility is that the `bonsai` dedispersion algorithm used by the realtime pipeline favors the ends of the bands. `bonsai` may not be uniformly sensitive across the CHIME bandpass to narrowband bursts, and could be less likely to detect events in the center of the band. Further analysis with system injections could be used to detect such an effect, but is beyond the scope of this paper. Such a selection effect would affect the distribution of peak frequencies for bursts in the published CHIME/FRB catalog (CHIME/FRB Collaboration et al. 2021), but this bimodality is not observed in any other CHIME/FRB repeating sources or in the population of one-off bursts.[2] The distribution of peak frequencies for all FRBs in the catalog does peak around 470 MHz, but this can be attributed to the increased beamwidths at lower frequencies.

---

[2] https://www.chime-frb.ca/catalog.

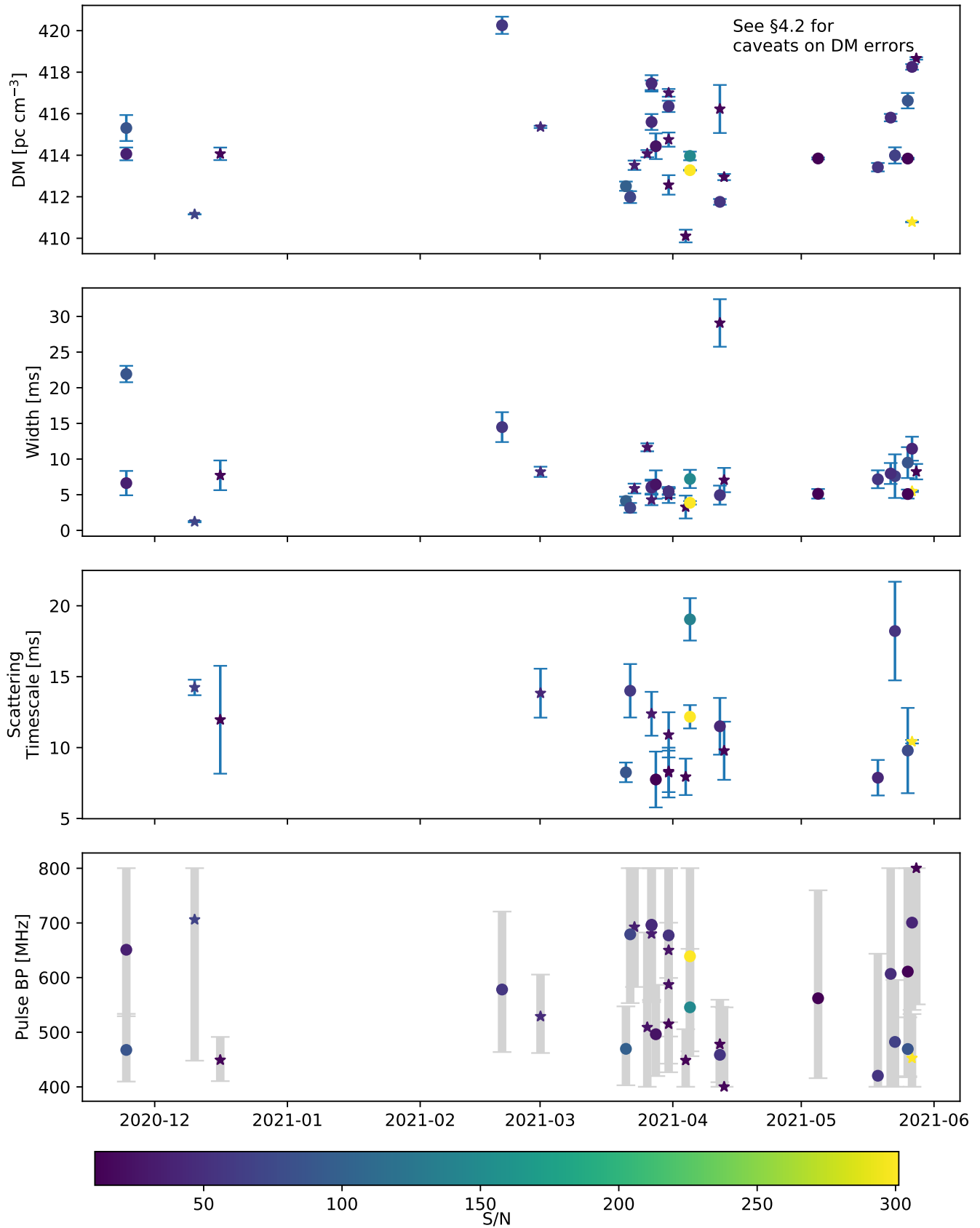

**Figure 4.** Measured properties of the bursts vs. time. The S/N of each burst, as reported by `fitburst`, is indicated with the point color. Events which fell outside the 600 MHz synthesized beamwidth (and hence whose fluence measurements are lower bounds only) are marked with stars. For the pulse bandpasses (BP), the dots indicate the location of the spectrum peak and the bars mark the bandpass.

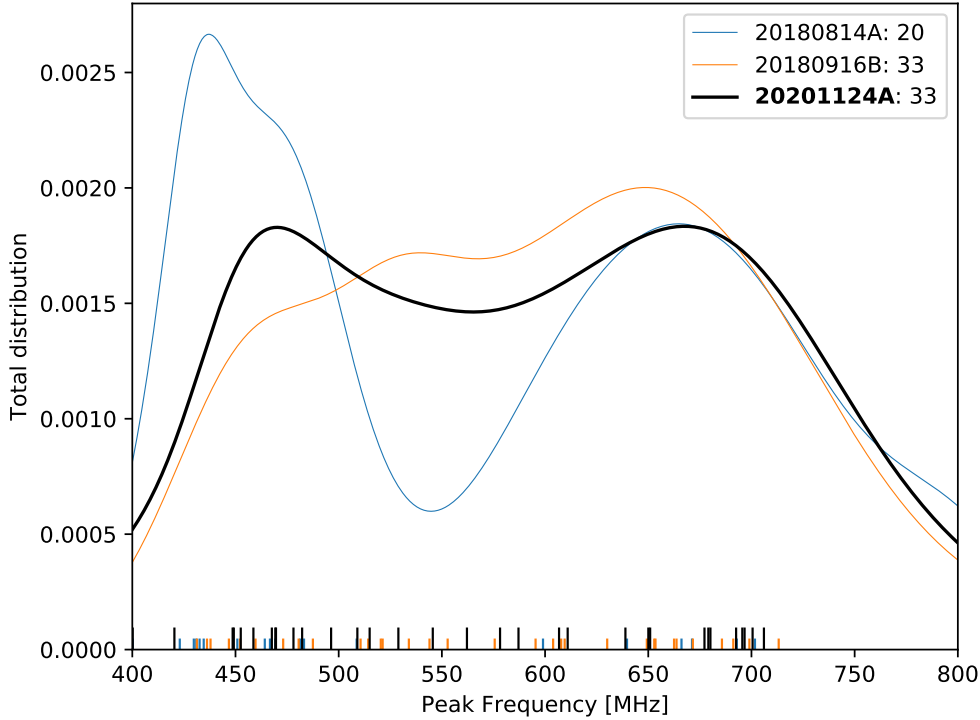

**Figure 5.** Total estimated distribution of peak frequencies for FRB 20201124A and two repeating FRBs from the first CHIME/FRB catalog (CHIME/FRB Collaboration et al. 2021) with the largest numbers of events, calculated using Equation (2). Marks on the bottom indicate locations of the peak frequencies for each distribution. The numbers in the legend indicate the total number of bursts from each. All three seem to peak around 668 MHz.

The bimodality could also come from complex chromatic sensitivity in the formed beams (Pleunis et al. 2021). A single burst could be observed to have different spectra in different beams depending on the source's sky location relative to the beam centers. The `fitburst` analysis here only fits the spectrum in the beam with the highest S/N, and 10 out of the 26 events reported here were also detected in one or two other beams. Fitting the spectrum in these other beams could in principle find parts of the spectrum missed by the highest S/N beam. We ran `fitburst` on these other beams, however, and found the bursts in the same part of the CHIME bandpass as in the highest S/N beams. Follow-up analysis with baseband data will let us beamform directly to the known source location, and avoid this chromaticity issue altogether. A full analysis of baseband data from CHIME/FRB repeaters will be the subject of future work.

It is difficult to make a meaningful comparison with detections in other bands. The uGMRT bursts shown in Marthi et al. (2021) peak around 700 MHz, but the total bandpass searched is too narrow to detect both peaks in Figure 5. Other instruments have detected tens to hundreds of narrowband bursts at GHz frequencies, but haven't reported distributions of peak frequencies.

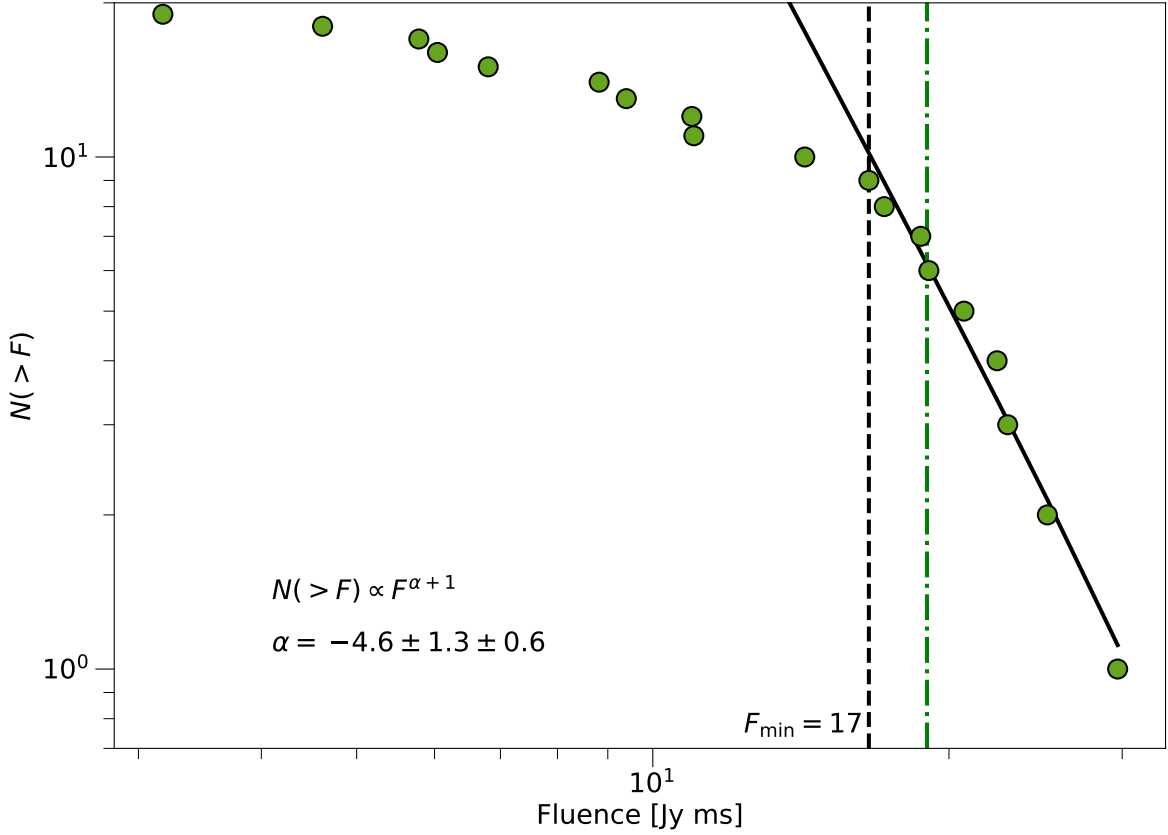

**Figure 6.** Fluence distribution of FRB 20201124A, with a maximum likelihood estimator fit for a power-law at high fluence and a low-fluence cutoff point. The dashed black line marks the fitted cutoff for the power-law, and the dashed green line marks the 90% fluence threshold. The two errors on $\alpha$ are the mean error of the MLE power-law fit and the standard deviation of fitted indices across the Monte-Carlo sampling of fluences.

### 4.3. *Luminosity Function*

CHIME/FRB Collaboration et al. (2020) reported the cumulative distribution of fluences for the periodic FRB 20180916B, fitting a power-law distribution $N(>F) \propto F^{\alpha+1}$ with a break at low fluence due to sample incompleteness. Using a maximum-likelihood estimator (MLE), they fitted a power-law index $\alpha = -2.3 \pm 0.3$ and a cutoff at $5.2 \pm 0.1$ Jy ms. This cutoff corresponded with a separate estimate of the fluence completeness limit using a Monte Carlo technique described by Josephy et al. (2019), which takes into account the effects of spectral energy distribution, source location along the beam, and daily gain variations.

We perform the same analysis here for FRB 20201124A, the results of which are shown in Figure 6. As in CHIME/FRB Collaboration et al. (2020), we first solve simultaneously for the power-law index and the turnover point $\hat{F}_{min}$ using the best-fit values of fluence (excluding events for which only a fluence lower bound could be found). The best-fit $\hat{F}_{min}$ is 17 Jy ms, marked by a vertical dashed black line in Figure 6, is in good agreement with the separately estimated 90% fluence completeness of 19 Jy ms. To account for the uncertainties in the measured fluences, we generate multiple realizations of the fluences by treating each as a Gaussian variable with FWHM given by the errors in Table 2. For each of 5000 realizations, we run the MLE power-law fit, holding $F_{min}$ fixed. The best-fit power-law index is $\alpha = -4.6 \pm 1.3 \pm 0.6$, where the first error is given by the mean of the MLE error across the ensemble and the second error is the standard deviation of the fitted spectral indices. This is much steeper than reported for previous repeaters (Wang et al. 2018; CHIME/FRB Collaboration et al. 2020), though given the large uncertainties, it is consistent with them (to $2\sigma$).

We note that FRBs 20201210A and 20210526C have fluence lower bounds of 25.7 Jy ms and 140.4 Jy ms, respectively, and are the only bursts detected with a fluence bound above the turnover point in Figure 6 that were not included in the fit due to being outside the synthesized beam. The synthesized beam weighting, used in calibrating the sources used in this fit, is typically a factor between 1 and 2. It is possible that if the true fluence of these two bursts were included in the fit, they would shift our fitted power-law index to a shallower value.

## 5. DISCUSSION & CONCLUSIONS

We have reported on the discovery of a new repeating FRB, 20201124A, previously announced by Lanman & the CHIME/FRB Collaboration (2021) following prolific activity. This activity was noted by other groups who have already localized the source (Day et al. 2021; Law et al. 2021; Xu et al. 2021) to a galaxy SDSS J050803.48+260338.0 at spectroscopic redshift $z = 0.0979 \pm 0.0001$ (Fong et al. 2021). We have presented the best-fit burst model parameters for 26 detected repeat bursts. Using the precise location provided by interferometric measurements (see Table 1), we have calculated the full exposure time history of CHIME/FRB to the source position and applied beam-weighting to get better fluence estimates.

Although CHIME/FRB only observes the location of FRB 20201124A for a limited time each day, the long and consistent history of CHIME's survey allows us to show that the activity of FRB 20201124A has changed significantly over time. FRB 20121102A and FRB 20180916B both show periodic clusters of activity, with periods of 157 and 16.35 days, respectively (Cruces et al. 2021; CHIME/FRB Collaboration et al. 2020). The increased activity of FRB 20201124A in April could be one such cluster in a similar periodic process. However, we find no evidence of long-term periodicity in its repetition. The event rate appears to have declined by the beginning of May (see Figure 1). Future work will follow up on any bursts observed after the last reported here (2021 May 27).

We observe an apparent bimodality of peak frequencies in pulse spectra. Some of the bursts occur low in the bandpass and are wider in time, while several are higher in the bandpass and are more compact in time. However, this apparent bimodality could be due to some yet undetermined selection effect in the CHIME/FRB pipeline; an analysis of baseband data — beyond the scope of this paper — may be helpful as it can avoid beam effects in the spectrum.

Some bursts show evidence of having more than one downward drifting component, and one burst (20201210A) can be fit well with two components. The degeneracy between multi-component structure and DM makes it difficult to get consistent fits for DM and scattering timescale. Taken at face value, the scattering times measured for the bursts (Table 2) suggest that detection of this source at LOFAR radio frequencies will be impossible. For a 10-ms scattering time, assuming a typical $-4$ scattering index, at 100 MHz the scattering time will be 13 s. On the other hand, a LOFAR detection would signal a far shorter scattering time in the CHIME band, and imply that the bursts we have observed have broad intrinsic structure that is misinterpreted here as scattering, a possibility given the relatively narrow spectral extent and the possibility of unresolved structure.

Given the known host spectroscopic redshift of $z = 0.0979 \pm 0.0001$ (Fong et al. 2021), the bursts reported on here have luminosities ranging from $7 \times 10^{39}$ to $2 \times 10^{41}$ erg s$^{-1}$, and energies in the range $9 \times 10^{37}$ to $3 \times 10^{39}$ erg. These are unremarkable, in the midrange of those observed for repeating FRBs. The fluence index we observe, $-4.6 \pm 1.3 \pm 0.6$, is consistent with that observed for the repeating FRB 20180916B ($-2.3 \pm 0.3 \pm 0.1$) in the same energy band (CHIME/FRB Collaboration et al. 2020).

Marthi et al. (2021), using uGMRT data, fitted a cumulative burst rate as a function of fluence of the form $R(>F) = 10$ hr$^{-1}(F/F_c)^{\alpha+1}$ with a power law index of $\alpha = -2.2 \pm 0.2$ and a completeness threshold of $F_c = 10$ Jy ms over 550 – 750 MHz. This is from a sample of 48 bursts from FRB 20201124A detected in a three hour observing window on 2021 April 05. Though their reported event rate is inconsistent with our analysis (see §4.1), their fluence distribution is consistent with our fitted $\alpha$ given our large uncertainties.

Fitting to the burst energy[3] distribution of FRB 20121102A, Wang et al. (2018) found a best-fit power-law index of $\alpha = -2.16 \pm 0.24$. More recently, Zhang et al. (2021) fitted a power-law with $\alpha = -1.86 \pm 0.02$ to the high-energy end of the luminosity distribution of FRB 20121102A (above $\sim 10^{38}$ erg) in the 1.4-GHz frequency range. Below that cutoff, the luminosity distribution deviates strongly from a power-law (Li et al. 2021).[4] The recent uGMRT observations of FRB 20201124A support the possibility of a universal luminosity law for repeaters, and our results do not contradict theirs. It will be interesting to see how luminosity functions from FAST observations of FRB 20201124A compare with our results.

The discovery of FRB 20201124A further underscores the power of sensitive wide-field radio telescopes like CHIME to continue to discover and report on repeating FRBs. Narrower-field radio telescopes then act as excellent follow-up instruments. Such complementary tag-team work is clearly highly efficient at pushing forward FRB science.

## 6. ACKNOWLEDGEMENTS

We are grateful to the staff of the Dominion Radio Astrophysical Observatory, which is operated by the National Research Council of Canada. CHIME is funded by a grant from the Canada Foundation for Innovation (CFI) 2012 Leading Edge Fund (Project 31170) and by contributions from the provinces of British Columbia, Québec and Ontario. The CHIME/FRB Project, which enabled development in common with the CHIME/Pulsar instrument, is funded by a grant from the CFI 2015 Innovation Fund (Project 33213) and by contributions from the provinces of British Columbia and Québec, and by the Dunlap Institute for Astronomy and Astrophysics at the University of Toronto. Additional support was provided by the Canadian Institute for Advanced Research (CIFAR), McGill University and the McGill Space Institute thanks to the Trottier Family Foundation, and the University of British Columbia.

FRB research at MIT is supported by NSF Grant 2008031. FRB research at UBC is supported by an NSERC Discovery Grant and by the Canadian Institute for Advanced Research. FRB research at WVU is supported by an NSF grant (2006548, 2018490) Research at the McGill Space Institute is supported by the Trottier Family Foundation. The CHIME/FRB Project is funded by a grant from the CFI 2015 Innovation Fund (Project 33213) and by contributions from the provinces of British Columbia and Quebec, and by the Dunlap Institute for Astronomy and Astrophysics at the University of Toronto. The Dunlap Institute is funded through an endowment established by the David Dunlap family and the University of Toronto. A.B.P. is a McGill Space Institute (MSI) Fellow and a Fonds de Recherche du Quebec – Nature et Technologies (FRQNT) postdoctoral fellow. B.M.G. is supported by an NSERC Discovery Grant (RGPIN-2015-05948), and by the Canada Research Chairs (CRC) program. C.L. was supported by the U.S. Department of Defense (DoD) through the National Defense Science & Engineering Graduate Fellowship (NDSEG) Program D.C.G. is supported by the John I. Watters Research Fellowship E.P. acknowledges funding from an NWO Veni Fellowship. Fengqiu (Adam) Dong is supported by the Four Year Fellowship K.S. is supported by the NSF Graduate Research Fellowship Program. M.B. is supported by an FRQNT Doctoral Research Award. P.C. is supported by an FRQNT Doctoral Research Award. P.S. is a Dunlap Fellow and an NSERC Postdoctoral Fellow. V.M.K. holds the Lorne Trottier Chair in Astrophysics & Cosmology and a Distinguished James McGill Professorship and receives support from an NSERC Discovery Grant and Herzberg Award, from an R. Howard Webster Foundation Fellowship from the Canadian Institute for Advanced Research (CIFAR), and from the FRQNT Centre de Recherche en Astrophysique du Quebec.

---

[3] Directly proportional to fluence.
[4] However, Aggarwal (2021) dispute the findings of Li et al. (2021), arguing that the FAST fluences were likely overestimated such that the part of the luminosity function showing non-power-law behavior should actually be below the completeness threshold of FAST.